\documentclass[preprint,pre]{revtex4}
\usepackage{amssymb}
\usepackage{amsmath}
\usepackage{graphicx}

\usepackage{xparse} % to draw a convex hull in RLT
\usepackage{listings} % JSON code highlighting
\usepackage{subcaption}
\usepackage{forest} % for RLT diagrams
\usetikzlibrary{shapes, tikzmark, positioning, patterns}
\usepackage{transparent} % for transparent framings
\usepackage{soul} % provides \hl{..}

\usepackage{url}
\usepackage[a4paper]{geometry}
\usepackage{hyperref}

\definecolor{colornumber}{HTML}{ccccff}
\definecolor{colorstring}{HTML}{fdae6b}
\definecolor{colorboolean}{HTML}{7fcdbb}

\newcommand{\convexpath}[2]{
  [   
  create hullcoords/.code={
    \global\edef\namelist{#1}
    \foreach [count=\counter] \nodename in \namelist {
      \global\edef\numberofnodes{\counter}
      \coordinate (hullcoord\counter) at (\nodename);
    }
    \coordinate (hullcoord0) at (hullcoord\numberofnodes);
    \pgfmathtruncatemacro\lastnumber{\numberofnodes+1}
    \coordinate (hullcoord\lastnumber) at (hullcoord1);
  },
  create hullcoords
  ]
  ($(hullcoord1)!#2!-90:(hullcoord0)$)
  \foreach [
  evaluate=\currentnode as \previousnode using \currentnode-1,
  evaluate=\currentnode as \nextnode using \currentnode+1
  ] \currentnode in {1,...,\numberofnodes} {
    let \p1 = ($(hullcoord\currentnode) - (hullcoord\previousnode)$),
    \n1 = {atan2(\y1,\x1) + 90},
    \p2 = ($(hullcoord\nextnode) - (hullcoord\currentnode)$),
    \n2 = {atan2(\y2,\x2) + 90},
    \n{delta} = {Mod(\n2-\n1,360) - 360}
    in 
    {arc [start angle=\n1, delta angle=\n{delta}, radius=#2]}
    -- ($(hullcoord\nextnode)!#2!-90:(hullcoord\currentnode)$) 
  }
}

\definecolor{eclipseStrings}{RGB}{42,0.0,255}
\definecolor{eclipseKeywords}{RGB}{127,0,85}
\colorlet{numb}{magenta!60!black}

\lstdefinelanguage{json}{
    basicstyle=\normalfont\ttfamily\linespread{0.01},
    commentstyle=\color{eclipseStrings}, % style of comment
    stringstyle=\color{eclipseKeywords}, % style of strings
    numbers=left,
    numberstyle=\scriptsize,
    stepnumber=1,
    numbersep=8pt,
    showstringspaces=false,
    breaklines=true,
    frame=left,
    rulecolor=\transparent{0},
    showlines=true, % show empty lines
    string=[s]{"}{"},
    comment=[l]{:\ "},
    morecomment=[l]{:"},
    literate=
        *{0}{{{\color{numb}0}}}{1}
         {1}{{{\color{numb}1}}}{1}
         {2}{{{\color{numb}2}}}{1}
         {3}{{{\color{numb}3}}}{1}
         {4}{{{\color{numb}4}}}{1}
         {5}{{{\color{numb}5}}}{1}
         {6}{{{\color{numb}6}}}{1}
         {7}{{{\color{numb}7}}}{1}
         {8}{{{\color{numb}8}}}{1}
         {9}{{{\color{numb}9}}}{1}
}

\ExplSyntaxOn
\NewDocumentCommand{\xsubfigure}{ m m }
 {% #1 is a symbolic key, #2 is a list of key-value pairs
  \roly_xsubfigure:nn { #1 } { #2 }
 }
\NewDocumentCommand{\makerow}{ m }
 {% #1 is a list of symbolic keys
  \roly_makerow:n { #1 }
 }

\keys_define:nn { roly/subfigures }
 {
  width .tl_set:N = \l_roly_subfig_width_tl,
  body .tl_set:N = \l_roly_subfig_body_tl,
  caption .tl_set:N = \l_roly_subfig_caption_tl,
 }

\dim_new:N \l_roly_row_height_dim
\box_new:N \l_roly_body_box

\cs_new_protected:Npn \roly_xsubfigure:nn #1 #2
 {
  \prop_if_exist:cTF { l_roly_subfig_#1_prop }
   {
    \prop_clear:c { l_roly_subfig_#1_prop }
   }
   {
    \prop_new:c { l_roly_subfig_#1_prop }
   }
  \keys_set:nn { roly/subfigures } { #2 }
  \prop_put:cnV { l_roly_subfig_#1_prop } { width } \l_roly_subfig_width_tl
  \prop_put:cnV { l_roly_subfig_#1_prop } { body } \l_roly_subfig_body_tl
  \prop_put:cnV { l_roly_subfig_#1_prop } { caption } \l_roly_subfig_caption_tl
 }

\cs_new_protected:Npn \roly_makerow:n #1
 {
  \dim_zero:N \l_roly_row_height_dim
  \clist_map_inline:nn { #1 }
   {
    \hbox_set:Nn \l_roly_body_box
     {
      \prop_item:cn { l_roly_subfig_##1_prop } { body }
     }
    \dim_compare:nT { \box_ht:N \l_roly_body_box > \l_roly_row_height_dim }
     {
      \dim_set:Nn \l_roly_row_height_dim { \box_ht:N \l_roly_body_box }
     }
   }
  \clist_map_inline:nn { #1 }
   {
    \begin{subfigure}[t]{ \prop_item:cn { l_roly_subfig_##1_prop } { width } }
    \raggedright
    \vspace{0pt} % for top alignment
    \parbox[t][\l_roly_row_height_dim]{\textwidth}{
      \prop_item:cn { l_roly_subfig_##1_prop } { body }
    }
    
    \caption{ \prop_item:cn { l_roly_subfig_##1_prop } { caption } }
    \end{subfigure}
    
   }
   \unskip\\ % end up the row
 }
\ExplSyntaxOff

\newtheorem{lem}{Lemma}

\begin{document}
\title{Tree edit distance for hierarchical data compatible with HMIL paradigm}
\author{B\v{r}etislav \v{S}op{\'i}k}
\email{bretislav.sopik@avast.com}
\author{Tom\'{a}\v{s} Stren\'{a}\v{c}ik}
\affiliation{Avast Software, Pikrtova 1737/1a, 14000 Prague, Czech Republic}

\maketitle

\section{Introduction}

In contemporary data analysis in industrial and academic research we often need to work with data that has a hierarchical structure. The analysis of such data is naturally more difficult than the analysis of data with a flat structure because the schema of the hierarchically organized dataset may possess important information which is lost if we ignore it. A common task of a dataset analysis is to evaluate the difference between two of its instances. If the dataset has a flat structure and consists of numerical vectors, appropriate distance function from vector spaces can be used. Similarly, we can utilize the Levenshtein edit distance\cite{Wagner-Fischer-1974} for comparison of two strings, etc. However, this task gets significantly more challenging when we consider hierarchically structured data consisting of fields of multiple data types. Because the schema of hierarchically organized data has a form of a rooted tree, the applicable distance function may belong to the class of \emph{tree edit distances} (TED). These distance functions are widely studied topic in the literature \cite{Bille-2005} and are a special case of more general edit distances on graphs \cite{Blumenthal-2019}.

There are many formats wich allow for the hierarchical organization of data, e.g., JSON\footnote{https://www.json.org/json-en.html}, XML\footnote{https://www.w3.org/XML/}, YAML\footnote{https://yaml.org/} or Apache Parquet\footnote{https://parquet.apache.org/}. In this work we focus on the JSON format which is arguably one of the most widely used. Interestingly, despite a significant body of research on TED, we are not aware of any implementation of TED applicable to hierarchically structured data in the JSON format apart of the approximate solution by Augsten \emph{et al}. \cite{Augsten-etal-2008-A, Augsten-etal-2008-B} which was implemented in a Python package \emph{jqgram}\footnote{https://github.com/hoonto/jqgram}. 

Nowadays, data analysis is a common part of a work on a machine learning (ML) problem. Solving ML problems on data with hierarchical structure is as challenging as the data analysis for the very same reason. Successful and generally applicable ML frameworks were missing, until recently, when Hierarchical multi-instance learning (HMIL) paradigm was developed.\cite{Pevny-Somol-2016, Mandlik-2021} It allows for end-to-end ML in AutoML fashion on hierarchically ordered data which contain structural elements in form of bags of equivalent observations (instances).

In this work, we present TED applicable to hierarchically structured data that is expressed in the JSON format and compatible with the HMIL framework. In practical terms the HMIL framework compatibility means that all JSON elements of the Array data type are interpreted as unordered bags. A more detailed discussion of the notion of HMIL compatibility can be found further in the text.

The manuscript is organized as follows: Section \ref{sec:theory} compares the JSON format with the HMIL framework and defines a mapping between them. It also outlines relevant parts of the TED research and explains why they are not applicable to our case. In section \ref{sec:edit-distance} we define the edit distance applicable to data in JSON format compatible with the HMIL framework. Section \ref{sec:algorithm} presents an algorithm for calculating the distance and proof that the distance is minimal. The concluding section \ref{sec:conclusion} reviews the proposed solution, presents a recipe for computation and discusses possible applications.

\section{Related theory}
\label{sec:theory}

\subsection{Comparison of JSON format and HMIL framework}

\subsubsection{JSON sample}

The JSON format recognizes elementary data types \emph{Boolean}, \emph{Number}, \emph{String} and \emph{null}. It also recognizes two composite data types \emph{Object} and \emph{Array}. \emph{Object} is a key-value dictionary in which the key is \emph{String} and value is any JSON data type. \emph{Array} is an ordered sequence of elements of any JSON data type. Empty \emph{Object} or \emph{Array} is allowed as well. A \emph{JSON sample} is a hierarchical structure consisting of elements of elementary or composite data types.

We define \emph{JSON schema} as an abstract representation of a JSON sample capturing its hierarchy, used data types of its structural elements and names of labels in Object type elements. Elaborate definition of the JSON schema is beyond the purpose of this work. A sufficient framework is the \emph{JSON Schema} project\footnote{https://json-schema.org/}. We say that a JSON sample $J$ obeys some schema $S$ if it is compatible with $S$ within the applied framework.

\subsubsection{HMIL sample}

As noted in the introduction, the HMIL paradigm was developed to allow for AutoML on data with hierarchical structure. This paradigm was implemented in a set of Julia libraries under a common name Hierarchical Multi-instance Learning Library (HMill)\footnote{https://github.com/CTUAvastLab/Mill.jl}. Let us describe the AutoML procedure within the HMill implementation: It begins by discovering the underlying schema of structural organization obeyed by all samples of the input dataset and represents it in \emph{HMill schema}. The HMill schema is then utilized to automatically create appropriate neural network model with precise topology and also to map samples of the input dataset into \emph{HMill samples}. Finally, the mapped HMill samples are used to train the model.

Every HMill sample has a structure of a tree composed of three types of abstract data nodes --- \emph{array node}, \emph{bag node} and \emph{product node}. In this tree, leaves are of type array node. Array nodes store information of elementary data types, categorical variables or numerical vectors and embed it into a relevant vector space. Inner nodes are formed by nodes of type bag node or product node and store the structural information. The bag node represents an unordered collection of elements of some type of the abstract data nodes. The product node represents an ordered tuple joining independent elements of abstract data node type.

For simplicity and clarity of the following discussion we define a notion of \emph{HMIL sample} which is an equivalent of the HMill sample with the only difference that the array node of the HMIL sample is just a data container and lacks the data embedding functionality. By doing this we restrain ourselves from technical discussions about the implementation of the embedding with no impact on the presented results of this work.

Similarly to the case of JSON samples, we can define a notion of HMIL schema. As a result we can also define a notion of HMIL sample or set of HMIL samples obeying a particular HMIL schema.

\subsubsection{Mapping of JSON sample into HMIL sample and HMIL compatibility}

In order to proceed further we define an isomorphism $\phi: \mathcal{J} \to \mathcal{H}$, where $\mathcal{J}$ is a set of all JSON samples and $\mathcal{H}$ is a set of all HMIL samples. It allows us to map any JSON sample $J \in \mathcal{J}$ into a HMIL sample $H = \phi(J)$, $H \in \mathcal{H}$. The key principles of the isomorphism $\phi$ are the following:
\begin{itemize}
 \item Field of elementary data type is mapped into the array node with the corresponding data type.
 \item Composite data type \emph{Object} is mapped into the product node. That is because the JSON \emph{Object} is a key-value dictionary which can be ordered by keys and every of its values is an independent JSON data type. Similarly, product node represents an ordered tuple of independent elements of abstract data node type.
 \item Composite data type \emph{Array} is mapped into the bag node. We treat every JSON \emph{Array} as an unordered collection of elements of a particular JSON data type. Thus it can be mapped into a bag node which represents an unordered collection of elements of particular abstract data node type.
\end{itemize}

The defined mapping of the JSON \emph{Array} type in $\phi$ has important consequences. Despite the JSON \emph{Array} is defined as an ordered sequence of elements, it may be utilized to contain elements where ordering is irrelevant. We argue that it is indeed commonly used this way. The defined isomorphism $\phi$ is correct for datasets where ordering in all its JSON \emph{Array} elements is irrelevant. We call such datasets \emph{HMIL compatible}. The proposed edit distance is correct for JSON samples from HMIL compatible datasets.

\subsection{Tree edit distances}
\label{subsec:tree-edit-distances}
The most commonly studied tree structure with respect to TED is a \emph{rooted labeled tree} (RLT) with labeled nodes and edit operations\cite{Bille-2005}:

\begin{itemize}
 \item \emph{Relabel}: Change the label of a node of the tree.
 \item \emph{Delete}: Delete a non-root node $v$ of the tree with parent $v'$. Connect all the children of $v$ with the parent $v'$.
 \item \emph{Insert}: Insert a node $v$ in the tree so that it has parent $v'$. Reconnect some of the children of $v'$ to be children of $v$.
\end{itemize}

This means that we are allowed to delete or insert inner nodes of the tree. However, other problem settings were considered as well, e.g., Selkow \cite{Selkow-1977} studied the edit distance problem where only deletions/insertions of the leaves or sub-trees were allowed.

Literature recognizes two types of RLT --- ordered and unordered. For ordered RLT every node of the tree has exact left-to-right ordering of its children. For unordered RLT none of the nodes has its children ordered. This property has huge impact on the algorithmic complexity of the edit distance computation. It was proven that for unordered RLT the computation of the edit distance is NP-hard problem \cite{Zhang-etal-1992}, whereas in the ordered case algorithms with polynomial asymptotic complexity were found \cite{Tai-1979, Zhang-Shasha-1989}.

\section{Edit distance}
\label{sec:edit-distance}

In order to define the edit distance $d_{\mathcal{J}}: \mathcal{J} \times \mathcal{J} \to \mathbb{R}_{\geq 0}$ on JSON samples of HMIL compatible dataset we need to find a way how to represent any HMIL sample as a RLT structure and also to define a suitable edit distance on such RLT structures. This in technical terms means that we aim to find appropriate isomorphism $\psi: \mathcal{H}_{\phi(\mathcal{J})} \to \mathcal{T}$ which maps $\mathcal{H}_{\phi(\mathcal{J})} = \lbrace\phi(J), \forall J \in \mathcal{J} \rbrace$, $\mathcal{H}_{\phi(\mathcal{J})} \subset \mathcal{H}$, into a set of all RLT structures $\mathcal{T}$. We also need to define edit operations applicable to the mapped RLT $\mathcal{T}_{\psi \circ \phi(\mathcal{J})} = \lbrace\psi(H), \forall H \in \mathcal{H}_{\phi(\mathcal{J})}\rbrace$, $\mathcal{T}_{\psi \circ \phi(\mathcal{J})} \subset \mathcal{T}$, which induce correct TED $d: \mathcal{T} \times \mathcal{T} \to \mathbb{R}_{\geq 0}$. Finally, the corresponding edit distance on JSON samples equals $d_{\mathcal{J}}(J_1, J_2) = d(\psi \circ \phi(J_1), \psi \circ \phi(J_2))$; $J_1, J_2 \in \mathcal{J}$.

We argue that none of the RLT structures studied in the literature (see subsection \ref{subsec:tree-edit-distances}) is a valid target $\mathcal{T}_{\psi \circ \phi(\mathcal{J})}$ in our use case for several reasons. First two reasons are connected to the RLT structures and the last one to the allowed edit operations:
\begin{enumerate}
 \item In general, HMIL compatible JSON sample can not be isomorphic to RLT which is exclusively ordered or unordered. The HMIL mapping maps JSON type \emph{Object} to ordered product node and \emph{Array} to unordered bag node. We must define TED on a tree structure which allows for nodes of both ordered and unordered type.
 \item HMIL compatible JSON sample is isomorphic to a HMIL sample in which the inner nodes store only the structural information and elementary information is stored in the array nodes. Similarly, the corresponding RLT must contain the elementary information only in its leaves.
 \item When evaluating TED between two HMIL compatible JSON samples which obey the same HMIL schema we want to be sure that any sequence of edit operations applied to them produces a sample which follows the HMIL schema. We call such edit operations \emph{schema-preserving} operations.
\end{enumerate}

Given the listed restrictions we define the following mapping of HMIL sample into RLT and schema-preserving edit operations which induce the appropriate TED.

\subsection{Mapping of HMIL sample into RLT}

\begin{figure}[h]
\newsavebox{\boxobject}
\newsavebox{\boxbag}
\newsavebox{\boxvaluenumber}
\newsavebox{\boxvaluestring}
\newsavebox{\boxvalueboolean}
\savebox{\boxobject}{\tikz{\node[circle, draw]{};}}
\savebox{\boxbag}{\tikz{\node[diamond, draw]{};}}
\savebox{\boxvaluenumber}{\tikz{\node[rectangle, draw, fill=colornumber, anchor=center]{\phantom{x}};}}
\savebox{\boxvaluestring}{\tikz{\node[rectangle, draw, fill=colorstring, anchor=center]{\phantom{x}};}}
\savebox{\boxvalueboolean}{\tikz{\node[rectangle, draw, fill=colorboolean, anchor=center]{\phantom{x}};}}

\begin{forest}
  object/.style={circle, draw, anchor=center},
  bag/.style={diamond, draw, anchor=center},
  valuenumber/.style={rectangle,draw, fill=colornumber, anchor=center},
  valuestring/.style={rectangle,draw, fill=colorstring, anchor=center},
  valueboolean/.style={rectangle,draw, fill=colorboolean, anchor=center}
  [{},object, 
    for tree={
      l sep=2cm, s sep+=0.9em, grow=south
    }
    [{\phantom{x}}, valuestring, edge label = {node [midway,fill=white] {$k_1$}}]
    [{},object, edge label = {node [midway, fill=white] {$k_2$}}
      [{\phantom{x}}, valueboolean, edge label = {node [midway, fill=white] {$k_5$}}]
      [{\phantom{x}}, valuenumber, edge label = {node [midway, fill=white] {$k_6$}}]
      [{}, object, edge label = {node [midway, fill=white] {$k_7$}}]]
    [{}, bag, edge label = {node [midway, fill=white] {$k_3$}}
      [{\phantom{x}}, valuestring]
      [{\phantom{x}}, valuestring]
      [{\phantom{x}}, valuestring]
      [{\phantom{x}}, valuestring]]
    [{}, bag, edge label = {node [midway, fill=white] {$k_4$}}]
  ]
  \node [anchor=north west] at (current bounding box.north east) {%
    \begin{tabular}{| c l |}
      \hline
      \usebox{\boxobject} & Object Node \\
      \usebox{\boxbag} & Bag Node \\
      \usebox{\boxvaluestring} & Value Node --- String \\
      \usebox{\boxvalueboolean} & Value Node --- Boolean \\
      \usebox{\boxvaluenumber} & Value Node --- Number \\
      \hline
    \end{tabular}};
\end{forest}
\caption{A Rooted Labeled Tree. The node type and node data type are represented by the node shape and color, respectively. Edges are labeled as $k_1,\dots,k_7$. Child nodes of a Bag node are indistinguishable by their edge labels.}\label{fig:rlt}
\end{figure}
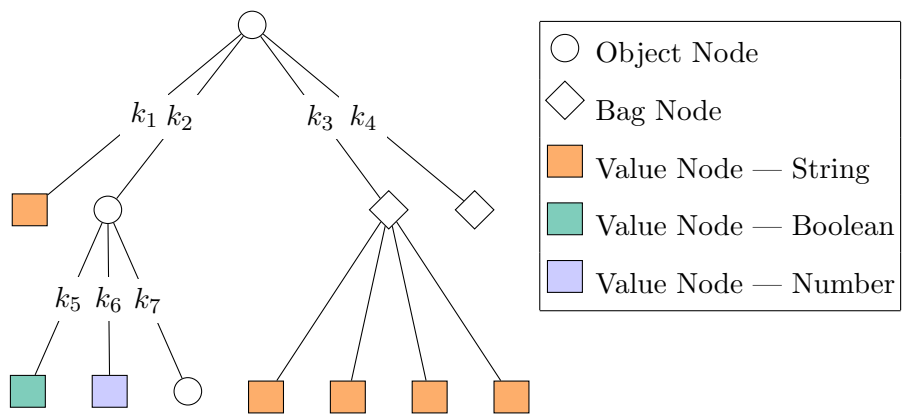

Let us define three \emph{node types} a tree $T$ may consist of: Bag, Object and Value. We define several functions acting on the nodes. Function $\nu(v)$ returns the \emph{node type} of a node $v$. Function $l(v)$ returns \emph{label of the node} $v$. Function $\chi(v)$ returns a \emph{bag of all children} of the node $v$. In addition, let us have two functions acting on the edges between nodes. Function $l(v \dashrightarrow w)$ returns \emph{label of the edge} $v \dashrightarrow w$ and function $\kappa(v)$ returns a \emph{bag of all labels of edges} between a node $v$ and its children $\chi(v)$. Finally, we also define \emph{data type} of the label of the node $v$ which is one of those supported by the JSON format: null, Boolean, String, Number. The data type of a label $l(v)$ is obtained by function $\tau(v)$. Figure {\ref{fig:rlt}} illustrates an arbitrary RLT structure.

We define isomorphic mapping $\psi$ of a HMIL sample $H \in \mathcal{H}$ into a corresponding RLT structure $T = \psi(H)$, $T \in \mathcal{T}$ as follows:

\begin{itemize}
 \item Product node in $H$ is mapped into $T$ as an Object node with labeled edges that are mutually unique. If $\nu(v) = \text{Object}$ then $|\chi(v)| = |\kappa(v)|$.  The labels of edges are equal to those in the corresponding JSON Object element.
 
 \item Bag node in $H$ is mapped into $T$ as a Bag node with unordered edges with no labels. If $\nu(v) = \text{Bag}$ then $\kappa(v) = \emptyset$. 

 \item Product node and bag node in $H$ are inner nodes of the HMIL schema, they describe only the hierarchical structure. This implies that the labels of the corresponding inner nodes in $T$ have implicitly a null value. If $\nu(v) \in \lbrace\text{Bag}, \text{Object}\rbrace$ then $l(v) = \text{null}$.
 
 \item Array node in $H$ is mapped into $T$ as a Value node with label equal to the data type and value of the corresponding field in the JSON sample. A node of type Value has no children, thus if $\nu(v) = \text{Value}$ then $\chi(v) = \emptyset$.

\end{itemize}

\begin{figure}[h!bt]
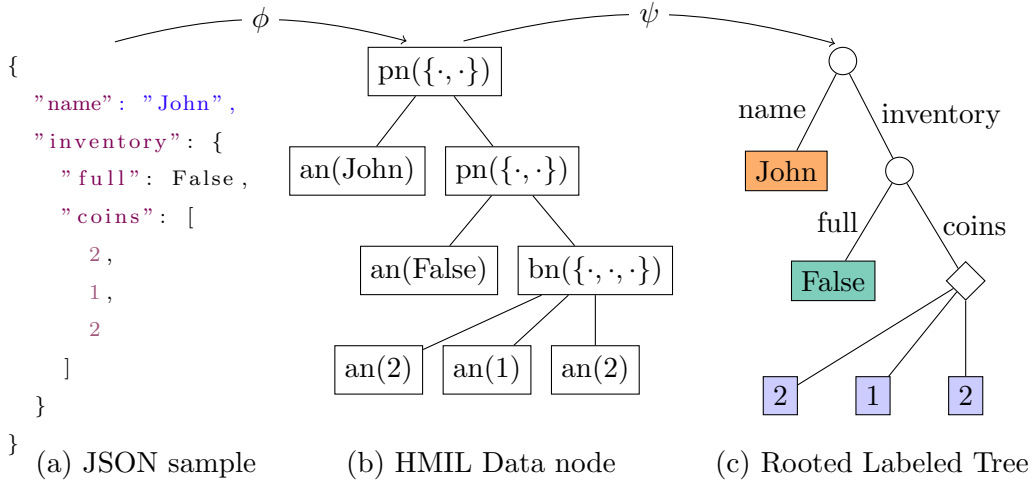

%%%%%%%%%%%%%%%%%%%%%%%%%%%%%%%%%%%%%%%%%%%%%%%%%%%%%%%%%%%%%%%%%%%%%%%
% This creates the isomorphism figure consisting of 3 subfigures.
% The special code is necessary to align both the captions of subfigures
% and also the images. Note the code subfigure is problematic and disrupts
% the alignment. For this reason it is styled with lstlisting with 
% transparent rectangular frame around code. Then it works.
\captionsetup[subfigure]{justification=justified,singlelinecheck=false}
\centering

\xsubfigure{sf:json}{
  width=0.25\textwidth,
  body={\lstinputlisting[language=json,numbers=none,escapechar=\%,basicstyle=\scriptsize]{figures/example.json}},
  caption={JSON sample}
}
\xsubfigure{sf:hmil}{
  width=0.35\textwidth,
  body={\input{figures/isomorphism_hmil}},
  caption={HMIL Data node}
}
\xsubfigure{sf:rlt}{
  width=0.35\textwidth,
  body={\input{figures/isomorphism_rlt}},
  caption={Rooted Labeled Tree}
}

\makerow{sf:json,sf:hmil,sf:rlt}

\begin{tikzpicture}[remember picture,overlay]
  \draw[->] ([yshift=0.4cm,xshift=1.25cm]pic cs:phi_l) to[bend left=15] node[midway, fill=white] {$\phi$} ([xshift=-0.5cm,yshift=0.5cm]pic cs:psi_l);
  \draw[->] ([yshift=0.5cm,xshift=0.25cm]pic cs:psi_l) to[bend left=15] node[midway, fill=white] {$\psi$} ([xshift=-3pt,yshift=0.6em]pic cs:psi_r);
\end{tikzpicture}
\caption{Three representations of arbitrary hierarchical sample. The mapping between (a) and (b) is done by $\phi$ and the mapping between (b) and (c) is realized by $\psi$. In (b), the \emph{an}, \emph{bn} and \emph{pn} represent the \emph{array}, \emph{bag} and \emph{product} nodes in accordance with the HMIL terminology.}\label{fig:isomorphism}
\end{figure}

Figure {\ref{fig:isomorphism}} depicts the mappings between a JSON sample, corresponding HMIL sample and a mapped RLT structure.

If the samples of the HMIL dataset obey some HMIL schema we utilize the isomorphism $\psi$ to define the corresponding \emph{RLT schema} which puts structural limitations on $\mathcal{T}_{\psi \circ \phi(\mathcal{J})}$. The schema-preserving edit operations are those which preserve the compatibility of $T$ with the corresponding RLT schema.

\subsection{Schema-preserving edit operations}

Edit operations which violate the RLT schema of $T$ --- and subsequently the HMIL schema of the corresponding $H$ --- are those that change its inner nodes. Thus, in order to define the schema-preserving edit operations we allow only those that act on the leaves of $T$.

We depict edit operations as $v \to v'$ where $v$, $v'$ are nodes of $T$. We also use a special symbol $\lambda$ to depict a non-existing node. The allowed edit operations read:
\begin{itemize}
 \item \emph{Relabel operation} $v \to v'$: We may change the label of a node $v$ of a type Value, $\nu(v) = \text{Value}$, to a value of the same data type $\tau(v) = \tau(v')$.
 \item \emph{Delete operation} $v \to \lambda$: We may delete a leaf $v$ of $T$, i.e., a node with $\chi(v) = \emptyset$, with label of any data type and value.
 \item \emph{Insert operation} $\lambda \to v$: We may insert a leaf $v$ into $T$ of any node type. In case the dataset follows some RLT schema the resulting tree must be compatible with it.
\end{itemize}

The defined elementary edit operations can be composed into a sequence of $k \in \mathbb{N}$ operations $S = s_1 \circ s_2 \circ \dots s_k$. By applying $S$ we transform a tree $T_1$ into $T_2$ which we depict as $T_1 \to T_2$. In this sequence, we begin with $T_1 = T_{(0)}$ and each edit operation $s_i$ performs transformation $T_{(i - 1)} \to T_{(i)}$ until we reach the final tree $T_2 = T_{(k)}$. As a corollary, the defined edit operations chain into sequences which allow only for deletion, insertion or relabeling of whole subtrees.

\subsection{Cost function}

The cost function $\gamma$ connected with relabeling, deleting or inserting a leaf of a tree may differ for every elementary type and actual compared values. We constraint $\gamma$ to be a proper distance function defined on nodes of the same type. Let us define a type function $\tau$ returning the data type of the node $v$. Then for any nodes $v_1, v_2, v_3$ of the same node type and data type the function $\gamma$ has the following properties:
\begin{align}
 \gamma(v_1, v_1) & = 0 && \text{identity of indiscernibles}, \\
 \gamma(v_1, v_2) & = \gamma(v_2, v_1) && \text{symmetry}, \\
 \gamma(v_1, v_2) & \leq \gamma(v_1, v_3) + \gamma(v_3, v_2) && \text{triangular inequality}, \\
 \text{which also implies} &&& \nonumber \\
 \gamma(v_1, v_2) & \geq 0 && \text{non-negativity}.
\end{align}
We extend $\gamma$ to the sequence of edit operations $S$ by $\gamma(S) = \sum_{i=1}^{|S|}\gamma(s_i)$, where the notation $\gamma(s_i)$ is a placeholder for $\gamma(v, v')$ which is cost of the edit operation $s_i$ involving nodes $v \to v'$. Finally, the distance function $d$ between trees $T_1, T_2 \in \mathcal{T}$ is defined as
\begin{equation} \label{distance_def_sequence}
 d(T_1, T_2) = \min_S \lbrace \gamma(S)| \text{$S$ is a sequence of edit operations transforming $T_1$ to $T_2$}  \rbrace.
\end{equation}
Since $\gamma$ is a distance function $d$ is a distance function too.

\subsection{Edit distance mapping}

Similarly to literature on TED\cite{Zhang-Shasha-1989, Bille-2005} we find it helpful to define \emph{edit distance mapping} of transformation $T_1 \to T_2$ denoted as $(M, T_1, T_2)$ (or $M$ for short). It is a set of tuples of nodes from $T_1$ and $T_2$ that are related via edit operations
\begin{equation}
 (M, T_1, T_2) = \lbrace (v, w) | \text{$w \in T_2$ is a transformation target of $v \in T_1$} \rbrace,
\end{equation}
Let us have a function $\pi(v)$ returning a \emph{parent of a node} $v \in T$ if exists. The defined edit operations imply that the mapping $M$ satisfies the following properties:
\begin{enumerate}
 \item Nodes in a tuple are of the same node type and data type, i.e., $\forall (v, w) \in M$: $\nu(v) = \nu(w)$ and $\tau(v) = \tau(w)$.
 \item The mapping is one-to-one, i.e., $\forall (v_1, w_1), (v_2, w_2) \in M$: $v_1 = v_2$ iff $w_1 = w_2$.
 \item Parents are part of the mapping as well, i.e., if $(v, w) \in M$ then $(\pi(v), \pi(w)) \in M$.
 \item If $(v, w) \notin M$ then no descendant of $v \in T_1$ and of $w \in T_2$ is in $M$.
 \item Labels of edges from parents are equal, i.e., $\forall (v, w) \in M$: if $\nu(\pi(v)) = \text{Object}$ then $l(\pi(v) \dashrightarrow v) = l(\pi(w) \dashrightarrow w)$.
\end{enumerate}
Let $N_1$ and $N_2$ be the set of nodes in $T_1$ and $T_2$, respectively, not included in $M$. We can define a \emph{cost of the mapping} $M$ as
\begin{equation} \label{mapping_cost}
 \gamma(M) = \sum_{(v, w) \in M} \gamma(v \to w) + \sum_{v \in N_1} \gamma(v \to \lambda) + \sum_{w \in N_2} \gamma(\lambda \to w).
\end{equation}
Following the proof in \cite{Zhang-Shasha-1989} we obtain the following relation between mapping and sequence of edit operations.
\begin{lem} \label{lemma_1}
If $(M, T_1, T_2)$ is a mapping of edit sequence $S$ transforming $T_1$ to $T_2$ then $\gamma(M) \leq \gamma(S)$. For any mapping $(M, T_1, T_2)$ there exists an edit sequence such that $\gamma(S) = \gamma(M)$. 
\end{lem}
By combination of \eqref{distance_def_sequence} and Lemma~\ref{lemma_1} we obtain that the minimum cost mapping is equivalent to the edit distance
\begin{equation}\label{distance_def_mapping}
 d(T_1, T_2) = \min_{M} \lbrace \gamma(M)|\text{$(M, T_1, T_2)$ is an edit distance mapping} \rbrace.
\end{equation}

\section{Algorithm for calculating the minimum cost}
\label{sec:algorithm}

We label an empty tree as $\Lambda$ and we define a function $\rho(T)$ which returns a \emph{root node} of a tree $T$. Let us have a tree $A$ with a root $a = \rho(A)$ and the rest of nodes divided into a set of $m \geq 0$ mutually disjoint subtrees $\lbrace A_1, \dots, A_m \rbrace$. Then we have the following recursive rules evaluating the costs of creating or deleting $A$:
\begin{align}
 d(\Lambda, \Lambda) & = 0, \\
 d(A, \Lambda) & = \sum_{i = 1}^{m}d(A_i, \Lambda) + \gamma(a, \lambda), \label{edit_distance_removal}\\
 d(\Lambda, A) & = \gamma(\lambda, a) + \sum_{i = 1}^{m}d(\Lambda, A_i). \label{edit_distance_insertion}
\end{align}

Let us also have a tree $B$ with $b = \rho(B)$ and set of $n \geq 0$ mutually disjoint subtrees $\lbrace B_1, \dots, B_n \rbrace$. We have:
\begin{itemize}
 \item If $\nu(a) \neq \nu(b)$ then
 \begin{equation} \label{edit_distance_nodes_neq}
  d(A, B) = d(A, \Lambda) + d(\Lambda, B).
 \end{equation}
 \item If $\nu(a) = \nu(b) \land \nu(a) = \text{Value}$ then
 \begin{equation} \label{edit_distance_nodes_value}
  d(A, B) =
   \begin{cases}
   \gamma(a, b)\quad\text{if $\tau(a) = \tau(b)$,} \\
   \gamma(a, \lambda) + \gamma(\lambda, b)\quad\text{if $\tau(a) \neq \tau(b)$}.
   \end{cases}
 \end{equation}
 \item If $\nu(a) = \nu(b) \land \nu(a) = \text{Object}$ then
 \begin{multline} \label{edit_distance_nodes_object}
  d(A, B) = \sum_{i = 1}^{m} \sum_{j = 1}^{n} d(A_i, B_j) \,\delta\left(l(a \dashrightarrow \rho(A_i)), l(b \dashrightarrow \rho(B_j))\right) \\
  + \sum_{k \in \kappa(a) \setminus \kappa(b)} \sum_{i = 1}^{m} d(A_i, \Lambda) \, \delta\left(l(a \dashrightarrow \rho(A_i)), k\right) \\
  + \sum_{k \in \kappa(b) \setminus \kappa(a)} \sum_{j = 1}^{n} d(\Lambda, B_j) \, \delta\left(l(b \dashrightarrow \rho(B_j)), k\right).
 \end{multline}
 Here the Kroenecker $\delta$-function reads
 \begin{equation}
  \delta(x, y) =
   \begin{cases}
    1\quad\text{if $x = y$,} \\
    0\quad\text{otherwise}.
   \end{cases}
 \end{equation}
 The first contribution in \eqref{edit_distance_nodes_object} is a sum of distances between subtrees $d(A_i, B_j)$ which share the same edge label $l(a \dashrightarrow \rho(A_i))$ and $l(b \dashrightarrow \rho(B_j))$, respectively. The second contribution is a sum of deletion costs $d(A_i, \Lambda)$ of subtrees $A_i$ which do not have edge label $l(a \dashrightarrow \rho(A_i))$ in $\kappa(b)$. Similarly, the third contribution is a sum of insertion costs $d(\Lambda, B_j)$ of subtrees $B_j$ which do not have edge label $l(b \dashrightarrow \rho(B_j))$ in $\kappa(a)$.
 \item If $\nu(a) = \nu(b) \land \nu(a) = \text{Bag}$ then
 \begin{equation} \label{edit_distance_nodes_bag}
  d(A, B) = \sum_{i,j = 1}^{\max(m, n)} \mathcal{M}_{ij} \Delta_{ij}.
 \end{equation}
 Here $\Delta$ is a square matrix of distances with $\text{rank}(\Delta) = \max(m,n)$ and elements
 \begin{equation}
  \Delta_{ij} =
  \begin{cases}
   d(A_i, B_j)\quad\text{for $1 \leq i \leq m$ and $1 \leq j \leq n$}; \\
   d(A_i, \Lambda)\quad\text{for $1 \leq i \leq m$ and $n < j \leq m$, if $m > n$}; \\
   d(\Lambda, B_j)\quad\text{for $m < i \leq n$ and $1 \leq j \leq n$, if $n > m$}.
  \end{cases}
 \end{equation}

  Matrix $\mathcal{M}$ is minimum cost mapping of $\Delta$ with elements $\mathcal{M}_{ij} \in \lbrace0, 1\rbrace$. The mapping can be found by application of, e.g., Hungarian\cite{Kuhn-1955} or BlossomV\cite{Kolmogorov-2009} algorithm.
  
\end{itemize}

An illustration of the edit distance calculation between two rooted labeled trees is included in Appendix {\ref{app:example}}.

\subsection{Proof of the minimum cost}

Our aim is to prove that equations \eqref{edit_distance_nodes_neq}-\eqref{edit_distance_nodes_object} and \eqref{edit_distance_nodes_bag} calculate costs of the minimum cost mapping $(M, A, B)$ and hence the edit distance $d(A, B)$ as defined in \eqref{distance_def_mapping}. The situation $\nu(a) \neq \nu(b)$ implies not only that $(a, b) \notin M$ but consequently that $M = \emptyset$. The equation \eqref{edit_distance_nodes_neq} is a direct application of \eqref{mapping_cost} to such case. The situation $\nu(a) = \nu(b) \land \nu(a) = \text{Value}$ leads to a mere evaluation of the difference of the particular Value nodes in \eqref{edit_distance_nodes_value}. The situation $\nu(a) = \nu(b) \land \nu(a) = \text{Object}$ is also straightforward because all nodes $(\rho(A_i), \rho(B_j))$ that share the same edge label $l(a \dashrightarrow \rho(A_i)) = l(b \dashrightarrow \rho(B_j))$ also belong to the mapping $(\rho(A_i), \rho(B_j)) \in M$. Again, the equation \eqref{edit_distance_nodes_object} is a direct application of \eqref{mapping_cost}. The last situation $\nu(a) = \nu(b) \land \nu(a) = \text{Bag}$ is resolved by calculating the cost according to the found minimum cost mapping of subtrees $\lbrace A_1, \dots, A_m \rbrace \cup \lbrace \Lambda \rbrace$ to $\lbrace B_1, \dots, B_n \rbrace \cup \lbrace \Lambda \rbrace$. We conclude that the calculated cost in each situation represents the cost of the minimum cost mapping and thus is equal to the distance $d(A, B)$.

\section{Conclusion}
\label{sec:conclusion}

In this work, we have developed an edit distance applicable to JSON samples compatible with the HMIL framework. By HMIL compatibility we mean that all elements of JSON Array data type are interpreted as unordered bags. The distance between two JSON samples $J_1$ and $J_2$ is calculated first by representing them as rooted labeled trees $T_1 = \psi \circ \phi(J_1)$ and $T_2 = \psi \circ \phi(J_2)$ and then by applying formulas depicted in \eqref{edit_distance_nodes_neq}-\eqref{edit_distance_nodes_object} and \eqref{edit_distance_nodes_bag} recursively.

The proposed edit distance is applicable to any task where exact evaluation of the difference between hierarchically organized and HMIL compatible data is needed. An example of the use case within the HMIL framework is to use this edit distance to gain insight into the distribution of the training data. This can help to adjust the sampling of the dataset so that a sufficient number of samples is selected into the training of the HMIL model also from the weakly populated regions of the dataset phase space.

\bibliography{tree-edit-distance.bib}{}
\bibliographystyle{ieeetr}

%\hl{Appendix}
\clearpage
\appendix
\section{Example of the edit distance calculation}
\label{app:example}
\begin{figure}[h]
% the edit operation figure
\begin{forest}
  object/.style={circle, draw, anchor=center},
  bag/.style={diamond, draw, anchor=center},
  valuenumber/.style={rectangle,draw, fill=colornumber, anchor=center},
  valuestring/.style={rectangle,draw, fill=colorstring, anchor=center},
  valueboolean/.style={rectangle,draw, fill=colorboolean, anchor=center}
  [{},object, name=t1,
    for tree={
      l sep=2cm, s sep+=0.9em, grow=south
    }
    [{\phantom{x}}, valueboolean, edge label = {node [midway,fill=white] {$k_1$}}]
    [{},object, edge label = {node [midway, fill=white] {$k_2$}}
      [{A\tikzmark{e_l}}, valuestring, edge label = {node [midway, fill=white] {$k_5$}}]
      [{\phantom{x}}, valuenumber, edge label = {node [midway, fill=white] {$k_6$}}]
      [{}, object, edge label = {node [midway, fill=white] {$k_7$}}]]
    [{}, bag, name=a1, edge label = {node [midway, fill=white] {$k_3$}}
      [{\phantom{x}}, valuenumber, name=a2]
      [{\phantom{x}}, valuenumber, name=a3]
      [{\phantom{x}}, valuenumber, name=a4]
      [{\phantom{x}}, valuenumber, name=a5]]
    [{}, bag, name=a6, edge label = {node [midway, fill=white] {$k_4$}}]
  ]
  \draw[black,dashed] \convexpath{a1, a6, a5, a2}{0.4cm};
  \node[right=0.4cm of a6] {removal};
  \node[left=0.5cm of t1] {$T_1$};
\end{forest}\hfill
\begin{forest}
  object/.style={circle, draw, anchor=center},
  bag/.style={diamond, draw, anchor=center},
  valuenumber/.style={rectangle,draw, fill=colornumber, anchor=center},
  valuestring/.style={rectangle,draw, fill=colorstring, anchor=center},
  valueboolean/.style={rectangle,draw, fill=colorboolean, anchor=center}
  [{},object, name=t2,
    for tree={
      l sep=2cm, s sep+=0.9em, grow=south
    }
    [{\phantom{x}}, valueboolean, edge label = {node [midway,fill=white] {$k_1$}}]
    [{},object, edge label = {node [midway, fill=white] {$k_2$}}
      [{B\tikzmark{e_r}}, valuestring, edge label = {node [midway, fill=white] {$k_5$}}]
      [{\phantom{x}}, valuenumber, edge label = {node [midway, fill=white] {$k_6$}}]
      [{}, object, edge label = {node [midway, fill=white] {$k_7$}}]]
    [{}, object, name=b1, edge label = {node [midway, fill=white] {$k_8$}}
      [{\phantom{x}}, valuenumber, name=b2, edge label={node [midway, fill=white] {$k_9$}}]]
  ]
  \draw[black,dashed] \convexpath{b1, b2}{0.4cm};
  \node[left=0.15 of b1] {addition};
  \node[left=0.5cm of t2] {$T_2$};
\end{forest}

\begin{tikzpicture}[remember picture,overlay]
\draw[->, dashed] ([xshift=2pt,yshift=5pt]pic cs:e_l) to[bend left] node[below, fill=white, xshift=3cm,yshift=-5pt] {value change} ([xshift=-10pt,yshift=5pt]pic cs:e_r);
\end{tikzpicture}
\caption{Tree edit distance mapping $(M,T_1,T_2)$. Dashed lines represent edit operations. The color and shape of a node represents the node data type and node type respectively as shown in Figure \ref{fig:rlt}.}\label{fig:edit_op}
\end{figure}
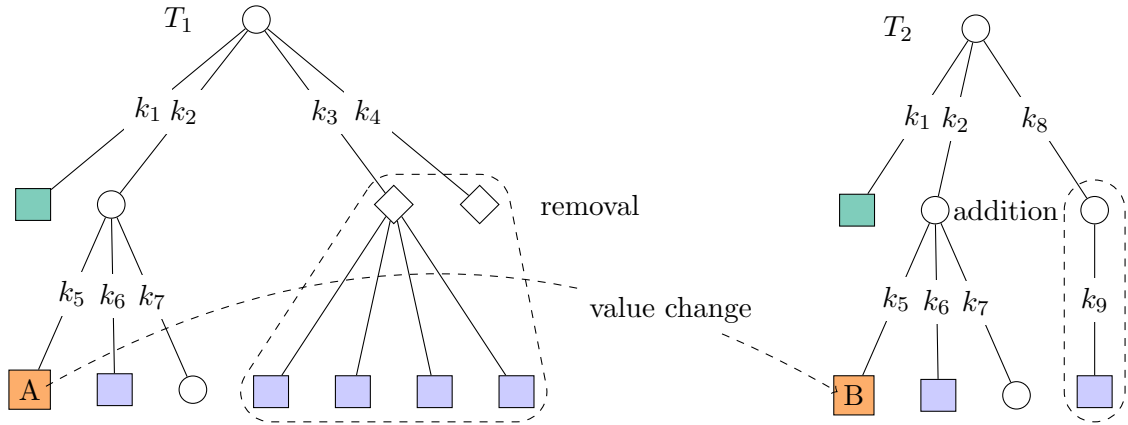

Let us now describe an example of the calculation of the edit distance between two RLT structures $T_1$ and $T_2$ as defined by formula \eqref{distance_def_mapping}. Figure \ref{fig:edit_op} captures the edit distance mapping $(M,T_1,T_2)$. For illustration purposes, we define the cost of edit operations for two nodes $u$ and $v$ of the same node type and data type (including the null node $\lambda$) to $\gamma(u,v)=1$. For simplicity, all operations of the edit sequence $S$ are depicted together in Figure \ref{fig:edit_op}. In this example, the edit sequence $S=S_1\circ S_2\circ S_3$ may be divided into three sequences of related edit operations:
\begin{itemize}
  \item{Node relabel operation ($S_1$): Change of a String Value node label in $T_1$ from value "A" to "B" in $T_2$. According to the equation \eqref{edit_distance_nodes_value}, the cost of this operation equals $\gamma($"A"$,$"B"$)=1$.}
  \item{Bag node removal ($S_2$): Two Bag nodes (and all their child nodes) are removed from $T_1$. In accordance with equation \eqref{edit_distance_removal}, this operation consists of the edit sequence $S_2=s_1\circ s_2\circ\dots\circ s_6$ with operations $s_1,\dots,s_4$ removing a Value node and $s_5,s_6$ removing a leaf Bag node. For each removed node $v$ the total edit distance increases by $\gamma(v,\lambda)=1$.}
  \item{Insertion of an Object node with one child node ($S_3$): The edit distance is increased by two, as described in equation \eqref{edit_distance_insertion}. First, an Object node with no children is inserted into $T_1$. Then, a Value node is inserted into the recently added Object node. The edge labels $k_8$ and $k_9$ are arbitrary in this case.}
\end{itemize}
As a result, the edit distance between $T_1$ and $T_2$ equals $d(T_1,T_2)=9$.

\end{document}